\documentclass[aps,12pt,superscriptaddress,amsfonts,amssymb,amsmath]{revtex4-2}
\pdfoutput=1

\usepackage{graphicx}
\usepackage{epsfig}
\usepackage{makeidx}
\usepackage{xcolor}
\usepackage{amsmath}
\usepackage{bm}
\usepackage{amssymb}

\begin{document}

\title{
Scalar-tensor gravity and Aharonov-Bohm electrodynamics with bosons: applications to superconductors
} 

\author{F.\ Minotti \footnote{Email address: minotti@df.uba.ar}}
\affiliation{Universidad de Buenos Aires, Facultad de Ciencias Exactas y Naturales, Departamento de F\'{\i}sica, Buenos Aires, Argentina}
\affiliation{CONICET-Universidad de Buenos Aires, Instituto de F\'{\i}sica Interdisciplinaria y Aplicada (INFINA), Buenos Aires, Argentina}

\author{G.\ Modanese \footnote{Email address: giovanni.modanese@unibz.it}}
\affiliation{Free University of Bozen-Bolzano \\ Faculty of Engineering \\ I-39100 Bolzano, Italy}
\date{\today}

\linespread{0.9}

\begin{abstract}

We study a scalar-tensor extension of gravity with two scalar fields coupled to the Aharonov-Bohm extension of electrodynamics, where the scalar mode $S\equiv\partial_\mu A^\mu$ is dynamical. In this framework the trace of the electromagnetic energy-momentum tensor is nonvanishing and the scalar $S$ induces an electro-gravitational coupling that can be enhanced by the vacuum expectation value of the second gravitational scalar. For bosonic matter described by a macroscopic wavefunction (as in superconductors), the coupling to the electromagnetic potential generates $S$ already at the semiclassical level, implying sizable junction-induced discontinuities. Including the scalar-tensor sector yields a nonlinear system for $S$ and a gravitational scalar combination $\beta$ that admits a bulk saturation solution $S_{\rm sat}^2=(\Lambda\lambda_L^2)^{-1}$ and a corresponding threshold condition for macroscopic effects. We apply these results to pulsed discharges across normal-superconducting junctions and obtain scaling relations for the onset of anomalous gravitational signals in terms of current density, pulse duration, and superconducting volume, consistent with reported threshold behavior in two independent experimental configurations for a single microscopic parameter. We also present time-dependent propagating solutions in the weak-field regime and derive a class of one-dimensional traveling exact solutions of the nonlinear vacuum Einstein equations.

\end{abstract}

\maketitle

\section{Introduction}

This paper investigates a coupled extension of gravity and electrodynamics obtained by combining a scalar--tensor theory of gravity with the Aharonov--Bohm extension of electrodynamics. On the gravitational side, the model contains the Brans--Dicke scalar field $\phi$ and a second scalar field $\psi$ \cite{minotti_modanese_mpla2025}. On the electromagnetic side, it adopts the Aharonov--Bohm formulation, in which the quantity
\begin{equation}
S=\partial_\mu A^\mu
\end{equation}
is promoted to a physical dynamical field. Within this framework, a direct coupling arises between the electromagnetic scalar $S$ and the gravitational sector, a feature that is absent in standard Maxwell electrodynamics.

Scalar--tensor extensions of General Relativity have long been studied in several contexts, including unification schemes, effective field theories, inflationary cosmology, and modified-gravity descriptions of dark energy and cosmic acceleration
\cite{Clifton2012ModifiedGravity,nojiri2017modified,buchbinder2017effective}. Their main appeal lies in the fact that they provide the simplest consistent enlargement of the gravitational field content beyond the metric tensor. The Aharonov--Bohm extension of electrodynamics has been explored much less extensively, but it is of particular interest because it allows the scalar $S$ to play a dynamical role and therefore permits couplings that are forbidden in Maxwell theory (\cite{Modanese2017MPLB,hively2019classical,minotti2023aharonov,EPJC2023} and refs.). In the present work, this feature becomes relevant in systems where the electromagnetic field interacts with a bosonic collective wavefunction.

Our matter system is a superconductor described by a macroscopic bosonic wavefunction $\psi_{bos}$. In particular, we consider pulsed high-current discharges through YBCO superconducting electrodes. YBCO is an extreme type-II superconductor, characterized by a coherence length $\xi$ much smaller than the magnetic penetration depth $\lambda_L$. For many purposes, the superconducting condensate can be described at leading order within the London approximation, in which $|\psi_{bos}|^2$ is treated as approximately constant in the bulk.

The main motivation for this study comes from two experimental claims of anomalous force generation in pulsed discharges involving YBCO electrodes \cite{PodkletnovModanese2001,Poher2011APR}. Although the two experimental configurations differ significantly in pulse duration, electrode thickness, and voltage scale, both reports indicate the presence of a threshold for the onset of the anomalous effect. The purpose of this paper is not to assess those experiments directly, but to examine whether the coupled scalar--tensor/electromagnetic model described above can account, at least phenomenologically, for the reported threshold behavior and for the order of magnitude of the inferred forces.

The basic mechanism can be summarized as follows. In the weak-field and quasi-static regime, the theory predicts that the Newtonian potential $\chi$, whose gradient determines the local gravitational acceleration, satisfies an equation of the form
\begin{equation}
\nabla^2 \chi=-\frac{1}{2}K_\phi c^2 S^2,
\label{nabla-chi}
\end{equation}
where $K_\phi$ parametrizes the effective coupling between the gravitational scalar sector and the electromagnetic scalar. Thus, whenever a sufficiently large $S$ field is generated, it acts as a source for an additional contribution to the local gravitational potential.

The generation of $S$ is especially effective in the bosonic case relevant to superconductors. For a collective wavefunction $\psi_{bos}$ coupled to the four-potential through the conserved Noether current, the field $S$ satisfies, in the absence of gravitational backreaction, the equation
\begin{equation}
\nabla^2 S=\frac{\mu_0 q}{m}\nabla\cdot\left(|\psi_{bos}|^2\mathbf{A}\right),
\end{equation}
which implies a discontinuity of $S$ across a normal--superconducting junction. The corresponding scale is set by the vector potential and the London penetration length, yielding a junction value of order
\begin{equation}
S_{jun}\sim \frac{A}{\lambda_L}.
\end{equation}
This is the first key ingredient of the mechanism.

The second key ingredient is the nonlinear backreaction of the gravitational scalars on the field $S$. Once the scalar gravitational coupling is included, the equation for $S$ takes the form
\begin{equation}
\Box[(1+\beta)S]=-\frac{\mu_0 q^2}{m}\nabla\cdot\left(|\psi_{bos}|^2\mathbf{A}\right),
\label{box-1-beta-S}
\end{equation}
where $\beta$ is a linear combination of the gravitational scalars $\phi$ and $\psi$, itself generated by $S$ through
\begin{equation}
\Box\beta=-\Lambda S^2.
\end{equation}
In the superconducting bulk, where $|\psi_{bos}|^2$ is approximately constant, these equations combine into a nonlinear equation for $S$ admitting a constant saturation solution,
\begin{equation}
S_{sat}^2=\frac{1}{\Lambda\lambda_L^2}.
\end{equation}
This means that, once the junction-generated value of $S$ reaches the appropriate level, the field does not simply decay over the London length, but can instead be sustained throughout the superconducting bulk. The onset of this regime defines a natural threshold condition for the appearance of anomalous gravitational effects.

A central result of the paper is that this threshold mechanism can be compared with the reported discharge experiments. Despite the differences between the two setups, the corresponding estimates lead to values of the vacuum expectation value $\psi_0$ of the second gravitational scalar that are of the same order of magnitude, namely $\psi_0\sim 10^{21}$. We regard this as an indication of internal consistency of the phenomenological framework, rather than as conclusive evidence for the model. The large value required for $\psi_0$ also suggests that the present Lagrangian should be viewed as an effective description valid in a restricted regime, and not necessarily as a complete theory applicable without modification at all scales.

Beyond the quasi-static analysis relevant to superconducting discharges, we also examine time-dependent traveling solutions of the coupled field equations. In the weak-field approximation, the scalar sector admits localized traveling configurations that can be associated with anomalous force propagation. We then discuss a class of one-dimensional exact vacuum traveling solutions of Einstein's equations, which illustrate the existence of nontrivial dynamical regimes in the theory. These results are exploratory in character and are presented mainly to clarify the range of behaviors allowed by the model.

Finally, we apply the same formalism to superconducting resonant cavities. In that case, for realistic parameters and for the value of $\psi_0$ inferred from pulsed discharges, the predicted anomalous forces are found to be far below present detectability. This comparison helps to delineate the parameter range in which the proposed mechanism may become relevant.

The paper is organized as follows. In Sect.~\ref{sec:fieldtheory} we summarize the underlying scalar--tensor and electromagnetic field equations in the weak-field approximation. In Sect.~\ref{pulsed-discharges} we analyze pulsed discharges in superconductors, derive the threshold condition, and compare the model with the discharge experiments. In the following sections we discuss time-dependent traveling solutions and their possible interpretation, and we then consider the case of superconducting cavities. Appendix~\ref{app2} is devoted to an internal consistency condition ensuring that the scalar couplings relevant here do not lead to observable deviations from ordinary gravitational physics in standard Solar-System conditions.

\section{The underlying field theory in short}
\label{sec:fieldtheory}

In this section we briefly review the field theory underlying the model. We give the general form of the action, the gravitational field equations in the weak-field approximation, and the electromagnetic field equations, distinguishing between coupling with fermionic wavefunctions and bosonic wavefunctions.

\subsection{The full action}

The general model we consider is described by the total action functional $S_{tot}$, in which we can distinguish the following terms: (a) a scalar-tensor Einstein gravitational action with a scalar field $\phi$ of the Brans-Dicke type; (b) a kinetic term for $\phi$; (c) the action of an additional external scalar field $\psi$ coupled to matter, to $\phi$ and to the e.m.\ fields through the scalar source $J$ (see below); (d) the matter action; (e) the action of the extended Aharonov-Bohm electrodynamics; (f) the coupling of the e.m.\ four-potential to the conserved electric four-current density $j^\mu_c$. All terms are covariantized using volume elements $\sqrt{-g}\, d\Omega$ and covariant derivatives $\nabla_{\mu}$. SI units are used. The full expression is

\begin{eqnarray}
S_{tot} &=&-\frac{c^{3}}{16\pi G_{0}}\int \sqrt{-g}\phi Rd\Omega +\frac{c^{3}}{
16\pi G_{0}}\int \sqrt{-g}\frac{\omega \left( \phi \right) }{\phi }\nabla
^{\nu }\phi \nabla _{\nu }\phi d\Omega  \nonumber \\
&&+\frac{c^{3}}{16\pi G_{0}}\int \sqrt{-g}\phi \left[ \frac{1}{2}\nabla
^{\nu }\psi \nabla _{\nu }\psi -U\left( \psi \right) -J\psi \right] d\Omega +
\frac{1}{c}\int \mathcal{L}_{mat}d\Omega  \nonumber \\
&&-\frac{\varepsilon _{0}c}{4}\int \sqrt{-g}\left[ F_{\mu \nu }F^{\mu \nu
}+2\left( \nabla ^{\mu }A_{\mu }\right) ^{2}\right] d\Omega -\frac{1}{c}\int 
\sqrt{-g}j^{\nu }_cA_{\nu }d\Omega .  \label{SKK}
\end{eqnarray}

The real scalar field $\psi$ can be regarded as the simplest way of giving an effective description of higher-dimensional theories and stabilizing the $\phi$ field (\cite{mbelek2003five,mbelek2004modelling} and refs.) Its Lagrangian density has the standard form above \cite{narlikar_book_1986} and the scalar source $J$ can be written as
\begin{eqnarray}
J &=&\beta _{mat}\left( \psi ,\phi \right) \frac{8\pi G_{0}}{c^{4}}\left(
T^{mat}+T^{AB}\right) +\beta _{AB}\left( \psi ,\phi \right) \frac{8\pi
G_{0}\varepsilon _{0}}{c^{2}}S^{2}  \nonumber \\
&&+\beta _{\phi }\left( \psi ,\phi \right) T^{\phi },  \label{ABsource}
\end{eqnarray}
where the $\beta ^{\prime }$s are non dimensional functions of the scalar
fields, $T^{mat}$ is the trace of the energy-momentum tensor of matter and $
T^{\phi }$ is the trace of the energy-momentum tensor of $\phi$. A remarkable feature of the Aharonov-Bohm electrodynamics is seen here, namely that it allows an interaction of the gravitational scalar $\psi$ with the e.m.\ scalar $S$, as well as with the trace $T^{AB}$ of the e.m.\ energy-momentum tensor, which is zero in Maxwell theory. The main interaction between $\psi$ and $S$, however, is mediated by the term proportional to $\beta_{AB}S^2$.

The scalar fields $\phi$ and $\psi$ have in general vacuum expectation values $\phi_0$ and $\psi_0$. In order to recover usual physics when the scalar fields are not excited, one must have $J(\phi_0,\psi_0)=0$. In the following, in order to simplify notations, we will denote by $\phi$ and $\psi$ their deviations with respect to the vacuum expectation values.

\subsection{Equations for the gravitational fields in weak-field approximation}
\label{sec:weak-field}

In our previous paper \cite{minotti_modanese_mpla2025}, the general field equations deriving from the action \eqref{SKK} were written and their weak-field approximation found, supposing that the deviations of the metric from the metric of flat spacetime are very small. For test particles in slow motion, the gravitational acceleration can be written as
\begin{equation}
    f_i=-\frac{\partial \chi}{\partial x^i}; \qquad i=1,2,3
\end{equation}
where $\chi$ is a Newtonian potential which receives contributions from matter sources in the usual way, plus contributions from the scalar fields described by the following equation:

\begin{equation}
\square \chi=\frac{\partial ^{2}\phi }{\partial t^{2}}+ \frac{1}{2c^2}K_\phi C_\phi S^{2}.
\label{potem}
\end{equation}

Here and in the following we use the constants 
\begin{subequations}
\label{CKs}
\begin{eqnarray}
K_{\phi } &=&\frac{8\pi G_{0}\varepsilon _{0}}{\left( 2\omega _{0}+3\right)
c^{2}}C_{\phi }, \\
K_{\psi } &=&\frac{8\pi G_{0}\varepsilon _{0}}{c^{2}}C_{\psi }, \\
C_{\phi } &=&\phi _{0}\psi _{0}\left. \frac{\partial \beta _{AB}}{\partial
\phi }\right\vert _{\phi _{0},\psi _{0}}, \\
C_{\psi } &=&\phi _{0}\psi _{0}\left. \frac{\partial \beta _{AB}}{\partial
\psi }\right\vert _{\phi _{0},\psi _{0}}.
\end{eqnarray}
\end{subequations}
The partial derivatives of the coupling $\beta_{AB}$ with respect to $\phi$ and $\psi$ evaluated at $\phi_0$ and $\psi_0$ account for the fact that this coupling depends on the scalars, but for small values of them only the linear dependence is relevant.

The equations for the gravitational scalars are
\begin{equation}
\square \phi=-K_\phi S^{2}.
\label{eq-phi}
\end{equation}
\begin{equation}
\square \psi=-K_\psi S^{2}.
\label{eq-phi}
\end{equation}

\subsection{Equations for the e.m.\ fields in weak-field approximation, with coupling to fermionic wavefunctions}

Concerning the e.m.\ part, when the electrical four-current density $j^\nu_c$ is of fermionic type (typically of the form $j^\nu_c=q\bar{\psi}_{ferm}\gamma^\nu \psi_{ferm}$), the equation for the e.m.\ potentials $A^\nu$ is
\begin{equation}
\square A^\nu=\mu_0 j^\nu_{c,ferm} -\partial^\nu(\beta S); \qquad\beta=C_\phi \phi+C_\psi \psi 
\label{em-potentials}
\end{equation}

Taking the four-divergence of this equation and remembering that, by definition, $S=\partial_\mu A^\mu$, one obtains the equation for $S$
\begin{equation}
\square [(1+\beta)S]=\mu_0 \partial_\nu j^\nu_{c,ferm} 
\label{eq-S}
\end{equation}
The four-divergence $\partial_\nu j^\nu_{c,ferm}$ is zero, classically. Therefore $S$ can be generated, in the fermionic case, only by possible quantum anomalies, as discussed in \cite{minotti2023aharonov,EPJC2023,minotti_modanese_math13050892}, resulting in small corrections to the Maxwell equations.

\subsection{Equations for the e.m.\ fields in weak-field approximation, with coupling to bosonic wavefunctions}
\label{sec:bosonic}

When the current is of bosonic type, the equation for the four-potentials is substantially different. We first write it without the gravitational term, as first given in our paper \cite{minotti_modanese_ijmpa2025}, split in the parts for the electric potential $\varphi=A^0/c$ and the vector potential $\mathbf{A}$:
\begin{equation}
\square \varphi=\frac{\rho}{\varepsilon_0} ,
\label{eq-pot}
\end{equation}
\begin{equation}
\square \mathbf{A}=\mu_0 \mathbf{j}_{c,bos}-\frac{\mu_0 q^2}{m}|\psi_{bos}|^2 \mathbf{A} .
\label{eq-A-vett}
\end{equation}

This equation is obtained from a Lagrangian in which the kinetic part coincides with the familiar Schr\"odinger Lagrangian, but the e.m.\ coupling is not obtained with the local gauge rule $\mathbf{p} \to \mathbf{p}+iq\mathbf{A}$; instead, it is obtained imposing a coupling of the form $A_\mu j_c^\mu$, where $j_c^\mu$ is the conserved N\"other current. 

(The two procedures give the same result in the fermionic case. All standard results concerning the Schr\"odinger equation for electrons are not affected.) 

As discussed in \cite{minotti_modanese_ijmpa2025}, we think the coupling $A_\mu j_c^\mu$ is the correct one, because it gives the correct correspondence between fermionic and bosonic currents. Since for bosonic wavefunctions it differs from the local gauge coupling, the consequence is that in the bosonic case one must renounce to full local gauge invariance and use for the pure e.m.\ part of the Lagrangian the Aharonov-Bohm extension of Maxwell theory.

From equations \eqref{eq-phi}, \eqref{eq-A-vett} we obtain, by taking the four-divergence, the equation for $S$
\begin{equation}
    \square S=-\frac{\mu_0 q}{m}\nabla\cdot (|\psi_{bos}|^2 \mathbf{A})
    \label{eq-for-S-non-grav}
\end{equation}
which is essential for establishing the existence of a discontinuity of $S$ at the superconducting junctions. The discontinuity is entirely due to the additional term proportional to $(|\psi_{bos}|^2 \mathbf{A})$ on the right-hand side\ of \eqref{eq-A-vett}.

After introducing the gravitational fields, eq.\ \eqref{eq-for-S-non-grav} becomes in weak field approximation the eq.\ \eqref{box-1-beta-S} mentioned in the Introduction and used in our main calculation in Sect.\ \ref{pulsed-discharges}, namely
\begin{equation}
  \Box [(1+\beta)S]=-\frac{\mu_0 q^2}{m} \nabla \cdot (|\psi_{bos}|^2 \mathbf{A})
\label{SMPLA}
\end{equation}

Finally we recall, concerning eq.\ \eqref{eq-A-vett}, that if $|\psi_{bos}|^2$ is spatially constant and equal to $n_s$ (the London approximation), the equation for $\mathbf{A}$ takes the form
\begin{equation}
  \Box \mathbf{A}+\lambda_L^{-2}\mathbf{A}=\mu_0 \mathbf{j}_{c,bos}
\end{equation}
implying a modification of the magnetic penetration length. We have discussed the compatibility of this modification with experimental data in the case of YBCO in \cite{minotti_modanese_ijmpa2025} and we are going to analyze this in  more detail in a forthcoming paper. In this paper, the London approximation is employed when we write the nonlinear equation for $S$ in the bulk of the superconductor (see eq.\ \eqref{Sscqs} and its constant solution in Sect.\ \ref{pulsed-discharges}).

\section{Pulsed discharges in superconductors}
\label{pulsed-discharges}

In this section we apply the model to the case of high-voltage electrical discharges in superconductors. We solve the linearized equations for the electromagnetic scalar field $S$ at junctions, where this field exhibits a discontinuity, and then the strongly nonlinear equation linking $S$ and the gravitational scalars in the bulk of the superconductor, and defining the threshold value of $S$. Finally, we relate this threshold value to the vector potential at the junction and to the current density. By inserting two different sets of experimental values for the currents, discharge times, superconductor volumes, and the observed anomalous gravitational forces, we obtain two very similar estimates for the parameter $\psi_0$. This is consistent with the model. The present calculation addresses the local source strength and threshold condition, not the full propagation problem of the far-field signal.

As detailed above, in the weak-field approximation of the theory in \cite{minotti_modanese_mpla2025}, the e.m.\ scalar $S$ in a superconductor, according to the theory developed in \cite{minotti_modanese_ijmpa2025}, is given by relation (\ref{SMPLA})
where, in terms of the magnitudes (\ref{CKs}), $\beta$ satisfies the equation
\begin{equation}
\square \beta =-\Lambda S^{2},  \label{Lapbeta}
\end{equation}
in which
\begin{eqnarray*}
\Lambda &=&C_{\phi }K_{\phi }+C_{\psi }K_{\psi }, \\
\beta &=&C_{\phi }\phi +C_{\psi }\psi .
\end{eqnarray*}

We can determine the jump of $\nabla S$ across a normal-superconductor
(N-SC) junction integrating eq.\ (\ref{SMPLA}) in a pill-box volume that includes the junction surface, of normal $\mathbf{n}$, to obtain
\begin{equation*}
\left( \left. \nabla \left[ \left( 1+\beta \right) S\right] \right\vert
_{+}-\left. \nabla \left[ \left( 1+\beta \right) S\right] \right\vert
_{-}\right) \cdot \mathbf{n}=-\frac{\mu _{0}q^{2}}{m}\left\vert \psi_{bos}
\right\vert ^{2}\mathbf{A}\cdot \mathbf{n}.
\end{equation*}
Since the integration of (\ref{Lapbeta}) in the same volume gives 
\begin{equation*}
\left( \left. \nabla \beta \right\vert _{+}-\left. \nabla \beta \right\vert
-\right) \cdot \mathbf{n}=0,
\end{equation*}
we finally have
\begin{equation}
\left( \left. \nabla S\right\vert _{+}-\left. \nabla S\right\vert
_{-}\right) \cdot \mathbf{n}=-\frac{\mu _{0}q^{2}}{m}\left\vert \psi_{bos}
\right\vert ^{2}\mathbf{A}\cdot \mathbf{n}.
\label{Dsjump}
\end{equation}
This jump occurs in a thin layer of thickness of order the coherence length $\xi$, in which the density of Copper pairs increases from zero to its value in the bulk. The value taken by the spatial derivative after the jump is expected to persist in a region of size close to the penetration length $\lambda _{L}$, which quantifies the possible spatial decay of $\mathbf{A}$.

Thus, if we have two parallel junctions, a N-SC at the left, and a SC-N a
distance $d$ to the right, in the first one there is a jump in $S$ of\ value
\begin{equation}
S\simeq -\frac{\mu _{0}q^{2}\lambda _{L}}{m}\left\vert \psi_{bos} \right\vert ^{2}
\mathbf{A}\cdot \mathbf{n}= -\frac{\mathbf{A}\cdot \mathbf{n}}{\lambda
_{L}}  \label{Sinside}
\end{equation}
with another one of similar magnitude, but of opposite sign, at the second
junction.

We further note that in the analyzed experiments the characteristic time of evolution of the discharge is long as compared with the time of transit of a signal at light speed in the superconducting region ($\mu$s as compared with ps). Thus, after a very rapid transient a quasi-stationary regime is established that follows the slow evolution of the current. 
There is a crucial point now. If we assume that the magnitude of $S$ is
small enough, so that, according to (\ref{Lapbeta}) the magnitude of $\beta $
is also small compared to 1, and can thus be neglected in (\ref{SMPLA}),
we then have from (\ref{SMPLA}) itself in the quasi-stationary regime that,
since $S=\nabla \cdot \mathbf{A}$, 
\begin{equation*}
\nabla ^{2}S=\frac{S}{\lambda _{L}^{2}},
\end{equation*}
which corresponds to a spatial exponential decay of $S$ with characteristic
length $\lambda _{L}$. In this way, $S$ is restricted to very thin layers
close to the junctions. If we determine the resultant force from the
gravitational potential
\begin{equation*}
\nabla ^{2}\chi =-\frac{K_{\phi }c^{2}}{2}S^{2},
\end{equation*}
we have
\begin{equation*}
\mathbf{f}=-\nabla \chi \sim K_{\phi }c^{2}S^{2}\lambda _{L}\sim K_{\phi
}c^{2}A^{2}\lambda _{L}^{-1}.
\end{equation*}

As is made clear below, in order to explain the forces reported by Poher and
Podkletnov using this expression, there are two difficulties. First, the
required magnitude of $K_{\phi }$ turns out to be too high to be compatible
with other experiments, for instance the superconducting resonant cavities (see Section \ref{cavities}). The second
problem is that Poher and Podkletnov results cannot be made compatible with
each other with a single value of $K_{\phi }$.

A resolution of this problem requires the inclusion of $\beta $, which
makes the problem non-linear. Expanding (\ref{SMPLA}) we have 
\begin{equation*}
\left( 1+\beta \right) \square S+S\square \beta +2\partial _{\nu }\beta
\partial ^{\nu }S=-\frac{\nabla \cdot \mathbf{A}}{\lambda _{L}^{2}},
\end{equation*}
and using (\ref{Lapbeta}) we obtain in the quasi-stationary regime
\begin{equation}
\left( 1+\beta \right) \nabla^{2} S+2\nabla\beta \cdot \nabla S=\frac{S}{\lambda _{L}^{2}}-\Lambda S^{3}.
\label{Sscqs}
\end{equation}

This relation has a solution with constant $S$ given by (note that all
involved magnitudes are positive definite)
\begin{equation}
S_{sat}^{2}=\frac{1}{\Lambda \lambda _{L}^{2}}.  \label{Ssat}
\end{equation}

We thus argue that the solution of $S$ self-adjusts to this value if
the material sources generating it are strong enough. In fact, the numerical solution of the 1-D version of eqs.\ (\ref{Lapbeta}) (its quasi-stationary version) and (\ref{Sscqs}), with the boundary conditions $\beta=0$, $d\beta/dx=0$, $S=0$, and $dS/dx$ non-zero, according to relation (\ref{Dsjump}), shows precisely that behavior for $S$, which takes on the constant value (\ref{Ssat}) after a growth in a thin region close to the boundary. This behavior is verified for any non-zero value of $dS/dx$ at the boundary, only that the smaller this derivative the larger is the region needed to reach the saturation value (\ref{Ssat}). In this way, there is a threshold below which $S$ cannot reach saturation within the superconductor.

In this solution the spatial decay associated to $\lambda _{L}$ is compensated by the non-linear effect of $\beta $, which is able to be effective only when $S$ reaches a high enough magnitude.

In this way, for the problem considered we have that, for strong enough
sources, a scalar is generated in a thin layer at one of the junctions until
it reaches the ``saturation'' value (\ref{Ssat}), stays constant in the bulk
of the superconductor, and decreases back to zero at the layer in the other
junction. If the superconductor is of thickness $d$, there is thus, right outside the superconductor, a specific force of magnitude
\begin{equation}
\mathbf{f}=-\nabla \chi \simeq \frac{K_{\phi }c^{2}}{\Lambda \lambda _{L}^{2}}
d= \frac{c^{2}}{C_{\phi }\lambda _{L}^{2}}d.  \label{funiv}
\end{equation}

To make further estimations we can evaluate $\mathbf{A}$, considering its
continuity across the junction, using the current density in the normal
medium (assumed to be cylindrical of radius $R$). We do not consider in this estimations the effect of magnetic flux penetration, which depends on several factors, but only a spatial average of  $\mathbf{A}$. For a uniform current
density $j$ we thus have ($\mathbf{n}$ is assumed in the $z$ direction)
\begin{equation*}
\mathbf{A}\cdot \mathbf{n}=A_{z}=-\frac{\mu _{0}j}{4}r^{2},
\end{equation*}
whose mean value in the junction area is 
\begin{equation*}
\overline{A}_{z}=-\frac{\mu _{0}j}{8}R^{2}=-\frac{\mu _{0}i_{c}}{8\pi },
\end{equation*}
with $i_{c}$ the total current. We thus have from (\ref{Sinside}) that a
scalar of magnitude 
\begin{equation*}
S\simeq \frac{\mu _{0}i_{c}}{8\pi \lambda _{L}}
\end{equation*}
can be generated at the junction.

In Poher's experiments values reported are $i_{c}\sim 6$ kA, which are
similar to those reported by Podkletnov, so that in both cases, taking $
\lambda _{L}\simeq 200$ nm, one would have scalars of magnitude $S\sim
2\times 10^{3}$ N A$^{-1}$m$^{-1}$. This
mechanism can account for the observed existence of a threshold for the effect to show up.

Poher, using a superconductor of total thickness $d\simeq 60$ $\mu $m and a discharge of
duration $16$ $\mu $s, reports $f\sim 1.5\times 10^{4}$ m s$^{-2}$. In
Podkletnov experiment, with $d$ of the order of 1 cm, and discharge durations of $200$
ns (determined from the reported capacitance, voltage and peak current), the reported specific impulse imparted to the test masses is about $1$ m s$^{-1}$, given a mean specific force $f\sim
5\times 10^{6}$ m s$^{-2}$. From (\ref{funiv}) we thus obtain (using again $\lambda _{L}\simeq 200$ nm) $C_{\phi }\simeq 9\times 10^{21}$ for Poher, and 
$C_{\phi }\simeq 4\times 10^{21}$ for Podkletnov, a good coincidence for two
experiments with so different parameters.

Since, considering a $\omega_{0}$ of order 1 (see discussion in Appendix \ref{app2}),
\begin{equation*}
\Lambda \simeq C_{\phi }K_{\phi }=\frac{8\pi G_{0}\varepsilon _{0}}{\left(
2\omega _{0}+3\right) c^{2}}C_{\phi }^{2}\sim 5\times 10^{5}\text{ A}^{2}
\text{N}^{-2},
\end{equation*}
we have from (\ref{Ssat}) $S_{c}\sim 7\times 10^{3}$ NA$^{-1}$m$^{-1}$, of
similar magnitude to the estimated scalar generated at the junctions in both
experiments. The corresponding value of $K_{\phi }$ is about $10^{-16\text{ }
}$A$^{2}$N$^{-2}$.

An interesting point is that the estimated value of the dimensionless
constant $C_{\phi }$ is of the same order of magnitude as the square root of
Dirac's big number (this is directly related to the question of $\psi $ VEV $
\psi _{0}$) 
\begin{equation*}
\frac{e^{2}}{4\pi \varepsilon _{0}G_{0}m^{2}}\simeq 4.1\times 10^{42}.
\end{equation*}

If we thus take $C_{\phi }^{2}$ to be given by this expression we have
\begin{equation*}
\Lambda \simeq \frac{8\pi G_{0}\varepsilon _{0}}{\left( 2\omega
_{0}+3\right) c^{2}}C_{\phi }^{2}=\frac{2}{\left( 2\omega _{0}+3\right) }
\left( \frac{e}{mc}\right) ^{2}.
\end{equation*}

Further noting that from (\ref{Ssat}) $\Lambda S_{c}^{2}\lambda _{L}^{2}=1$,
and that, according to (\ref{Sinside}), the generated scalar at the junction
is $S\sim A\lambda _{L}^{-1}$, we have that the condition for a sufficiently
strong source to generate the non-linear effect, $S\sim S_{c}$, corresponds
to $\Lambda A^{2}\sim 1$, that is
\begin{equation*}
\frac{eA}{mc}\sim 1,
\end{equation*}
which indicates that the condition is that the electromagnetic component of
the electron canonical momentum is of the same order of the mechanical
momentum $mc$.

\section{Time-dependent solutions}
\label{approximate-time-dep}

In this section we consider localized solutions with cylindrical symmetry, traveling at arbitrary velocities. For this, we resort to the weak-field approximation. We compare the obtained solutions with the Alcubierre ones, computing the corresponding metric in the covariant four-dimensional formalism and also in the (1+3)-dimensional formalism in terms of the shift vector and the lapse function. We note that while the Alcubierre solutions require a negative energy density source, in our case, the energy-momentum tensor of the scalar field can serve as the source. Finally, by moving from the weak-field approximation to the complete nonlinear gravitational equations in the vacuum, first with a $\phi$-field and then without it (pure Einstein equations), we derive a new class of one-dimensional traveling exact solutions.

One of the most baffling aspects in Podkletnov's pulsed experiments is the very high collimation and apparent superluminality of the pulses. Within the realm of existing theories we can argue that those characteristics might be the expression of a dynamical spacetime effect, analogous to that
considered by Alcubierre \cite{alcubierre1994} and others \cite{Lobo_Visser_2004,alcubierre_lobo_2017,Bobrick_2021}.

In this scenario a strong deviation of the spacetime metric, relative to that of Minkowski's, evolves as a very localized region moving for distant observers with arbitrary velocity $v$.

In the so called ``warp drive" spacetime solutions explored in the literature this region of spacetime is sustained by a comoving source of exotic matter that violates all or some of the energy conditions.

In this section we explore two possible scenarios for a dynamic spacetime. One of them is similar to that considered by Alcubierre, and in which the role of the exotic matter is played by the gravitational scalar $\phi$. In this scenario we must include the whole system of equations for the e.m.\ and gravitational scalars, together with Einstein equations for the metric of spacetime. Given the extreme complexity of the problem we will study only the weak-field approximation of those equations. This is a limitation because the sought effect involves a strong deviation of the metric. To proceed we must assume that this strong deviation occurs only in a thin region which is the boundary between the ``internal" solution, and the external one with Minkowski metric. The weak-field approximation is thus assumed valid in the internal region outside this boundary, and localized solutions traveling with arbitrary velocity $v$ are explored. 

The alternative scenario takes into account that the equations for the scalars are of the inhomogeneous wave-equation type, and thus considers that it is unlikely that superluminal propagation of those scalars can take place in vacuum. This particular aspect is explored with simple examples. For this reason we analyze the case in which the gravitational scalar cannot act as commoving source of the spacetime metric outside the superconductor. Consequently, we study the full, non-linearized Einstein equations in vacuum, and obtain exact 1-D traveling solutions with arbitrary velocity $v$.    

\subsection{Weak-field approximation (WFA)}

In order to work out analytical solutions we will further assume that the
different fields are functions of the variable 
\begin{equation*}
\xi =\sqrt{(z-vt)^{2}+r^{2}},
\end{equation*}
where $z$ is the coordinate corresponding to the symmetry axis, and along
which the pulse travels at arbitrary speed $v$, while $r$ determines the distance to this axis.

For a generic function $F\left( \xi \right) $ we thus have (the primes denote $\xi $\ derivatives)
\begin{eqnarray}
\square F &=&\frac{1}{c^{2}}\frac{\partial ^{2}F}{\partial t^{2}}-\frac{1}{r}
\frac{\partial }{\partial r}\left( r\frac{\partial F}{\partial r}\right) -
\frac{\partial ^{2}F}{\partial z^{2}}  \notag \\
&=&\left[ \lambda^{2}\left( 1-\frac{r^{2}}{\xi ^{2}}\right) -1\right]
\left( F^{\prime \prime }-\frac{F^{\prime }}{\xi }\right) +\left( \lambda^{2}-3\right) \frac{F^{\prime }}{\xi },  \label{DalaF}
\end{eqnarray}
where we have defined $\lambda=v/c$.

The WFA equations for the scalars of the theory to be satisfied outside the
superconductor are eq.\ (\ref{SMPLA}) without sources, and eq. (\ref{Lapbeta}):
\begin{subequations}
\label{WFAeqs}
\begin{eqnarray}
\square \left[ \left( 1+\beta \right) S\right] &=&0,  \label{WFA1} \\
\square \left[ \left( 1+\beta \right) \right] &=&-\Lambda S^{2},
\label{WFA2}
\end{eqnarray}
where we recall that $S$ is the e.m.\ scalar, $\beta =C_{\phi }\phi +C_{\psi }\psi $, where $
\phi $ and $\psi $ are the deviations of the scalars of the gravitation
theory, relative to their VEVs, and $\Lambda =C_{\phi }K_{\phi }+C_{\psi
}K_{\psi }$.

When working out the analytical solutions of the (\ref{WFAeqs}) we recall that these WFA equations are not expected to be valid in
regions with strong fields and/or field derivatives, so that we will consider piecewise
solutions valid inside some region only, and assume their match to the
solution outside that region to occur in a thin layer, across which either
the function and/or its spatial derivative can be discontinuous. Let thus
consider the solutions to be valid inside a region of size $L$ around the
center of the pulse located at $z_{c}=vt$, that is, for $\xi <L$.

If we consider that $\beta $ and $S$ are functions of only $\xi $, equation (\ref{WFA1}) can be written as, using (\ref{DalaF}) and (\ref{WFA2}), 
\end{subequations}
\begin{multline*}
\left( 1+\beta \right) \left\{ \left[ \lambda^{2}\left( 1-\frac{r^{2}
}{\xi ^{2}}\right) -1\right] \left( S^{\prime \prime }-\frac{S^{\prime }}{
\xi }\right) +\left( \lambda^{2}-3\right) \frac{S^{\prime }}{\xi }
\right\} \\
+2\beta ^{\prime }S^{\prime }\left[ \lambda^{2}\left( 1-\frac{r^{2}}{
\xi ^{2}}\right) -1\right] -\Lambda S^{3}=0.
\end{multline*}

By separating the terms with and without explicit $r$ dependence we obtain
the two separate equations 
\begin{subequations}
\label{sep1}
\begin{eqnarray}
\left( 1+\beta \right) \left( S^{\prime \prime }-\frac{S^{\prime }}{\xi }
\right) +2\beta ^{\prime }S^{\prime } &=&0, \\
\left( 3-\lambda^{2}\right) \left( 1+\beta \right) \frac{S^{\prime }
}{\xi }+\Lambda S^{3} &=&0.
\end{eqnarray}

We note from (\ref{sep1}) that for the particular case $\lambda^{2}=3$ the only possible solution is $S=0$ and $\beta$ constant. This indicates that non-trivial solutions depending on the single variable $\xi$ not always exist. In what follows we consider that $\lambda^{2}>3$.

The first of eqs.\ (\ref{sep1}) can be written as 
\end{subequations}
\begin{equation*}
\frac{2\beta ^{\prime }}{1+\beta }=\frac{1}{\xi }-\frac{S^{\prime \prime }}{
S^{\prime }},
\end{equation*}
which is immediately integrated to give
\begin{equation*}
\left( 1+\beta \right) ^{2}=\kappa \frac{\xi }{S^{\prime }},
\end{equation*}
with $\kappa $ an arbitrary constant.

Using the second of (\ref{sep1}) we finally have the relation between
functions
\begin{equation}
\frac{\left( 3-\lambda^{2}\right) \kappa }{1+\beta }=-\Lambda S^{3}.
\label{Sbeta}
\end{equation}

With (\ref{Sbeta}) the (\ref{WFA2}) is written as
\begin{equation*}
\square \left( \frac{1}{S^{3}}\right) =\frac{\Lambda ^{2}}{\left( 3-\lambda^{2}\right) \kappa }S^{2},
\end{equation*}
which in expanded form is
\begin{equation*}
\left( \lambda^{2}-1\right) \left( 1-\frac{r^{2}}{\xi ^{2}}\right) 
\left[ \frac{12S^{\prime ^{2}}}{S^{5}}-\frac{3}{S^{4}}\left( S^{\prime
\prime }-\frac{S^{\prime }}{\xi }\right) \right] +\frac{9S^{\prime }}{
S^{4}\xi }=\frac{\Lambda ^{2}}{\left( 3-\lambda^{2}\right) \kappa }
S^{2}.
\end{equation*}

By separating the terms with and without explicit $r$ dependence, we obtain
the pair of equations 
\begin{subequations}
\label{Scomp}
\begin{eqnarray}
S^{\prime \prime } &=&\frac{4S^{\prime 2}}{S}+\frac{S^{\prime }}{\xi }, \\
S^{\prime } &=&\Gamma \xi S^{6},
\end{eqnarray}
where 
\end{subequations}
\begin{equation}
\Gamma =\frac{\Lambda ^{2}}{9\left( 3-\lambda^{2}\right) \kappa }.
\label{Gamma}
\end{equation}

The (\ref{Scomp}) are two equations for the same function $S$ so that we
must study their compatibility. By derivation of the second of (\ref{Scomp})
we have (using the same equation again to replace the $S^{\prime }$
resulting in the derivation) 
\begin{equation*}
S^{\prime \prime }=\Gamma S^{6}+6\Gamma ^{2}\xi ^{2}S^{11}.
\end{equation*}

On the other hand, by replacing the second of (\ref{Scomp}) in the first one
we have
\begin{equation*}
S^{\prime \prime }=\Gamma S^{6}+4\Gamma ^{2}\xi ^{2}S^{11}.
\end{equation*}

We thus see that both (\ref{Scomp}) result in similar solutions if the
magnitude of $\Gamma ^{2}\xi ^{2}S^{11}$ is small compared to that of $
\Gamma S^{6}$, that is, if 
\begin{equation}
\left\vert \Gamma S^{5}\xi ^{2}\right\vert \ll 1.  \label{xicondition}
\end{equation}

Since our solutions are assumed to be valid in the region $\xi <L$, we must
require that $\left\vert \Gamma L^{2}S^{5}\right\vert \ll 1$.

With this condition we can determine $S$ using the second of (\ref{Scomp}),
which is easily integrated to give, with the arbitrary constant $S_{0}$,
\begin{equation*}
S=\frac{S_{0}}{\left( 1-\frac{5}{2}\Gamma S_{0}^{5}\xi ^{2}\right) ^{1/5}},
\end{equation*}
and, from (\ref{Sbeta}),
\begin{equation*}
1+\beta =-\frac{\Lambda }{9\Gamma S_{0}^{3}}\left( 1-\frac{5}{2}\Gamma
S_{0}^{5}\xi ^{2}\right) ^{3/5}.
\end{equation*}

Using the condition (\ref{xicondition}) we finally get our approximate
traveling solutions to be 
\begin{subequations}
\label{Travelfin}
\begin{eqnarray}
S &=&S_{0}\left( 1+\frac{1}{2}\Gamma S_{0}^{5}\xi ^{2}\right) ,  \label{Sfin}
\\
1+\beta &=&-\frac{\Lambda }{9\Gamma S_{0}^{3}}\left( 1-\frac{3}{2}\Gamma
S_{0}^{5}\xi ^{2}\right) .  \label{betafin}
\end{eqnarray}
Recalling that $\Gamma $ is given by (\ref{Gamma}), these solutions contain
the arbitrary constants $\kappa $ and $S_{0}$.

One important property of these solutions is that they satisfy the relation 
\end{subequations}
\begin{equation*}
F^{\prime \prime }-\frac{F^{\prime }}{\xi }=0,
\end{equation*}
so that, from (\ref{DalaF}), their D'Alembertians are simply: $\square
F=\left(\lambda^{2}-3\right) F^{\prime }/\xi $, which is a constant.

To determine the spacetime properties of the considered solutions we recall
the WFA equations obtained in \cite{minotti_modanese_mpla2025} for the metric tensor $g_{\mu \nu
}=\eta _{\mu \nu }+h_{\mu \nu }$, with $\eta _{\mu \nu }$ the metric of
Minkowski spacetime, with signature ($1$,$-1$,$-1$,$-1$). Greek indices
denote the four spacetime coordinates, with $x^{0}=ct$, while Latin indices
correspond to the three spatial coordinates.

The equations for the perturbation $h_{\mu \nu }$ are
\begin{equation}
R_{\mu \nu }-\frac{1}{2}Rg_{\mu \nu }\simeq -\square \overline{h}_{\mu \nu
}=2\left( \partial _{\mu \nu }\phi -\eta _{\mu \nu }\square \phi \right) ,
\label{hmn}
\end{equation}
where
\begin{equation}
\overline{h}_{\mu \nu }=h_{\mu \nu }-\frac{1}{2}\eta _{\mu \nu }\eta
^{\gamma \delta }h_{\gamma \delta },  \label{hbarmn}
\end{equation}
with the gauge 
\begin{equation}
\eta ^{\theta \mu }\partial _{\theta }\overline{h}_{\mu \nu }=0,
\label{gauge}
\end{equation}
which, since $\eta ^{\theta \mu }\partial _{\theta }\left( \partial _{\mu
\nu }\phi -\eta _{\mu \nu }\square \phi \right) $ is identically zero, can
always be satisfied by adding to $\overline{h}_{\mu \nu }$ any function $
\Theta _{\mu \nu }$ that satisfies $\square \Theta _{\mu \nu }=0$.

Considering the magnitude estimations in  \cite{minotti_modanese_mpla2025} we have that $\left\vert
C_{\phi }\right\vert \gg \left\vert C_{\psi }\right\vert $, which allows to
approximate $\phi \simeq $ $\beta /C_{\phi }$, so that $\phi $ is a
quadratic function of $\xi $, satisfying $\square \phi =\left(
\lambda^{2}-3\right) \phi ^{\prime }/\xi $.

In order to solve (\ref{hmn}) we write
\begin{equation}
\overline{h}_{\mu \nu }=H_{\mu \nu }\left( \xi \right) +\Theta _{\mu \nu },
\label{hbarsol}
\end{equation}
with $H_{\mu \nu }$ quadratic functions of $\xi $, and with $\square \Theta
_{\mu \nu }=0$.

In this way, we have that 
\begin{equation*}
\square \overline{h}_{\mu \nu }=\left( \lambda^{2}-3\right) \frac{
H_{\mu \nu }^{\prime }}{\xi }.
\end{equation*}

The evaluation of the right-hand side of (\ref{hmn}) is direct, resulting in
expressions proportional to $\phi ^{\prime }/\xi $ with different constants,
so that the integration is very simple, resulting in 
\begin{subequations}
\label{Hij}
\begin{eqnarray}
H_{00} &=&-\frac{6}{\lambda^{2}-3}\phi , \\
H_{03} &=&\lambda\frac{2}{\lambda^{2}-3}\phi , \\
H_{ij} &=&-2\frac{\lambda^{2}-2}{\lambda^{2}-3}\phi \delta
_{ij,}
\end{eqnarray}
with all other components equal to zero.

As convenient expressions satisfying $\square \Theta _{\mu \nu }=0$ we take 
\end{subequations}
\begin{equation*}
\Theta _{\mu \nu }=C_{\mu \nu }\left[ \xi ^{2}+\frac{1}{2}\left( \lambda^{2}-3\right) r^{2}\right] ,
\end{equation*}
with constants $C_{\mu \nu }$.

Since $C_{\mu \nu }=C_{\nu \mu }$ and, by the axial symmetry, $C_{11}=C_{22}$ and $C_{13}=C_{23}$, there are eight independent constants $C_{\mu \nu }$.
The gauge equations (\ref{gauge}) result in a simple algebraic system for
the constants $C_{\mu \nu }$, leading to the five independent equations
(note that $\phi ^{\prime }/\xi =\phi ^{\prime \prime }$ is a constant,
which we denote $\phi _{0}^{\prime \prime }$) 
\begin{subequations}
\label{Cij}
\begin{gather}
C_{10}=C_{20}=0, \\
C_{30}+\lambda C_{00}=\lambda\frac{2\phi _{0}^{\prime \prime }}{\lambda^{2}-3}, \\
C_{11}+C_{12}+C_{13}=\frac{2\left( \lambda^{2}-2\right) \phi
_{0}^{\prime \prime }}{\left( \lambda^{2}-3\right) \left( \lambda^{2}-1\right) }, \\
2C_{13}+C_{33}+\lambda C_{03}=-\frac{2\phi _{0}^{\prime \prime }}{\lambda^{2}-3}.
\end{gather}

Apart from the indeterminacy of three of the constants $C_{\mu \nu }$, each $
\overline{h}_{\mu \nu }$ is determined up to an additional constant. All these constants can
only be determined from the initial and boundary conditions of the full solution.

To more easily interpret the obtained solutions we express them in the $3+1$
formalism of general relativity that describes spacetime as a foliation of
space-like hypersurfaces at constant coordinate time $t$. In this
description the metric of spacetime is given as 
\end{subequations}
\begin{equation*}
ds^{2}=g_{\mu \nu }dx^{\mu }dx^{\nu }=\alpha ^{2}c^{2}dt^{2}-\gamma
_{ij}\left( dx^{i}+\tau ^{i}cdt\right) \left( dx^{j}+\tau ^{j}cdt\right) ,
\end{equation*}
where $\gamma _{ij}$ describes the 3-metric of the hypersurfaces, the shift
vector $\tau ^{i}$ relates the spatial coordinates of different
hypersurfaces, and $\alpha $ is the lapse function that gives the interval
of proper time between nearby hypersurfaces for observers with 4-velocity
normal to the hypersurfaces (Eulerian observers).

In the WFA used, in which $g_{\mu \nu }=\eta _{\mu \nu }+h_{\mu \nu }$,
using also relation (\ref{hbarmn}), it is easy to show that at first order
in the perturbations of the metric one has
\begin{eqnarray*}
\alpha &=&1-\frac{1}{4}\left( \overline{h}_{kk}+\overline{h}_{00}\right) , \\
\tau ^{i} &=&-\overline{h}_{0i}, \\
\gamma _{ij} &=&\delta _{ij}-\overline{h}_{ij}+\frac{1}{2}\left( \overline{h}
_{kk}-\overline{h}_{00}\right)\delta _{ij} .
\end{eqnarray*}

From the solutions (\ref{Hij}) and (\ref{Cij}) we have
\begin{eqnarray*}
\overline{h}_{kk}-\overline{h}_{00} &=&-6\phi +\Theta _{kk}-\Theta _{00}, \\
\overline{h}_{kk}+\overline{h}_{00} &=&-6\phi \frac{\lambda^{2}-1}{
\lambda^{2}-3}+\Theta _{kk}+\Theta _{00}, \\
\overline{h}_{01} &=&\overline{h}_{02}=0, \\
\overline{h}_{03} &=&\lambda\frac{2\phi }{\lambda^{2}-3}+\Theta
_{03}, \\
\overline{h}_{ij} &=&-2\frac{\lambda^{2}-2}{\lambda^{2}-3}
\phi \delta _{ij}+\Theta _{ij}.
\end{eqnarray*}

For comparison with Alcubierre's spacetime, in his case the geometry of
spacetime is given by 
\begin{eqnarray*}
\alpha &=&1, \\
\tau ^{1} &=&\tau ^{2}=0, \\
\gamma _{ij} &=&\delta _{ij},
\end{eqnarray*}
and with a $\tau ^{3}$ that differs from zero in a localized region moving with $
v $ along $x^{3}$.

In our solution we also have $\tau ^{1}=\tau ^{2}=0$, and a localized,
traveling at $v$, non-zero $\tau ^{3}$. However, the lapse function $\alpha $
and the 3-metric $\gamma _{ij}$ are perturbed about Alcubierre's values.

A peculiarity of Alcubierre's metric is that it violates the energy
conditions, requiring in particular that $R_{00}-\frac{1}{2}Rg_{00}<0$.
Related to this we see from (\ref{hmn}) that the scalar $\phi $ gives rise
to an effective energy-momentum tensor of the form
\begin{equation*}
T_{\mu \nu }^{eff}=\frac{c^{4}}{8\pi G_{0}}\left( \partial _{\mu \nu }\phi
-\eta _{\mu \nu }\square \phi \right) .
\end{equation*}

Since in the WFA the scalar $\phi $ satisfies the equation $\square \phi
=-K_{\phi }S^{2}$, from the obtained solution we have 
\begin{equation*}
T_{00}^{eff}=-\frac{3c^{4}}{8\pi G_{0}}\frac{K_{\phi }S^{2}}{\lambda^{2}-3},
\end{equation*}
which is negative if $K_{\phi }>0$ (we always consider in this section that $\lambda^{2}>3$).

Finally, we consider how action on test masses can be exerted in this case. For slowly moving neutral masses a specific gravitational force (per unit of mass) acting on them can be defined in the WFA in terms of the metric as \cite{minotti_modanese_mpla2025}
\begin{equation*}
f_{i}=c^{2}\left[ \frac{\partial \overline{h}_{0i}}{\partial x^{0}}-\frac{1}{
4}\frac{\partial }{\partial x^{i}}\left( \overline{h}_{00}+\overline{h}
_{kk}\right) \right] ,
\end{equation*}
which, from the equations of the theory in the WFA, can be expressed in
terms of a Newton-like gravitational potential $\chi $\ as
\begin{equation*}
f_{i}=-\frac{\partial \chi }{\partial x^{i}},
\end{equation*}
with $\chi $ satisfying the equation
\begin{equation*}
\square \chi =c^{2}\left( \frac{\partial ^{2}\phi }{\partial x^{02}}-\frac{1
}{2}\square \phi \right) .
\end{equation*}

From the obtained solution we have 
\begin{equation*}
\chi ^{\prime }\left( \xi \right) =\frac{c^{2}}{2}\frac{\lambda^{2}+3
}{\lambda^{2}-3}\phi ^{\prime }\left( \xi \right) ,
\end{equation*}
and the corresponding force along the axis
\begin{eqnarray*}
f_{z} &=&-\frac{c^{2}}{2}\left( z-vt\right) \frac{\lambda^{2}+3}{
\lambda^{2}-3}\frac{\phi ^{\prime }\left( \xi \right) }{\xi } \\
&=&\frac{c^{2}}{2}\left( z-vt\right) \frac{\lambda^{2}+3}{\left( 
\lambda^{2}-3\right) ^{2}}K_{\phi }S^{2}.
\end{eqnarray*}
This expression, valid inside the bubble, shows a linear dependence on the distance to the bubble center, being zero at that point. For $K_{\phi}>0$ the force is thus positive ahead, and negative behind, the bubble center.

\subsection{1-D traveling solutions}

In order to study an alternative possibility of a dynamic spacetime we consider in this section the dynamics of a simple wave equation in one spatial dimension, as representative of the equations for the scalars of the theory, and contrast it with the dynamics of the highly non-linear equations resulting when the metric of spacetime is not linearized around its Minkowski values. 

\subsubsection{Wave equation with superluminal sources}

We recall that the existence of effectively superluminal sources is well within the realm of Relativity theory \cite{Fayngold_book}. A familiar example is the commonplace reflection of light at a surface, in which the reflected light front corresponds to the light Mach cone (or wedge) generated by the superluminally moving excitation of the surface molecules by the incident light front. 

Let us then consider a 1-D wave equation of the form
\begin{equation}
\frac{1}{c^{2}}\frac{\partial ^{2}\Psi }{\partial t^{2}}-\frac{\partial
^{2}\Psi }{\partial x^{2}}=F\left( x-vt\right) ,  \label{wavesuper}
\end{equation}
where the velocity $v$ is arbitrary, and the source $F$ is a localized
function. The general solution can be written as
\begin{equation}
\Psi \left( x,t\right) =\Psi _{p}\left( x,t\right) +f\left( x-ct\right)
+g\left( x+ct\right) ,  \label{Psigen}
\end{equation}
where $\Psi _{p}\left( x,t\right) $ is a particular solution of (\ref
{wavesuper}).

To put a concrete example that can be treated analytically we take $
F=F_{0}\sec $h$^{2}\left( k\xi \right) $, with $\xi =$ $x-vt$.

In this way, looking for a $\Psi _{p}\left( \xi \right) $ we have from (\ref
{wavesuper}), with the primes indicating derivative with respect to the
argument of the function, (we recall the definition $\lambda=v/c$)
\begin{equation*}
\left( \lambda^{2}-1\right) \Psi _{p}^{\prime \prime }\left( \xi
\right) =F_{0}\text{ sech}^{2}\left( k\xi \right) ,
\end{equation*}
which can be integrated at once to give, with $c_{1,2}$ constants,
\begin{equation*}
\Psi _{p}\left( \xi \right) =\frac{F_{0}}{k^{2}\left( \lambda^{2}-1\right) }\ln \left[ \cosh \left( k\xi \right) \right] +c_{1}\xi +c_{2}.
\end{equation*}

If we impose as initial conditions that 
\begin{equation}
\Psi \left( x,0\right) =\left. \frac{\partial \Psi }{\partial t}\right\vert
_{t=0}=0,  \label{initialc}
\end{equation}
we have from (\ref{Psigen})
\begin{eqnarray*}
f\left( x\right) +g\left( x\right) &=&-\frac{F_{0}}{k^{2}\left( \lambda^{2}-1\right) }\ln \left[ \cosh \left( kx\right) \right] -c_{1}x-c_{2}, \\
f^{\prime }\left( x\right) -g^{\prime }\left( x\right) &=&-\frac{F_{0}}{
k\left(\lambda^{2}-1\right) }\tanh \left( kx\right) -c_{1}\lambda.
\end{eqnarray*}

Using the derivative of the first equation, together with the second one we
easily obtain
\begin{eqnarray*}
f^{\prime }\left( x\right) &=&-\frac{1}{2}\left( 1+\lambda\right) \left[ 
\frac{F_{0}}{k\left( \lambda^{2}-1\right) }\tanh \left( kx\right)
+c_{1}\right] , \\
g^{\prime }\left( x\right) &=&-\frac{1}{2}\left( 1-\lambda\right) \left[ 
\frac{F_{0}}{k\left( \lambda^{2}-1\right) }\tanh \left( kx\right)
+c_{1}\right] ,
\end{eqnarray*}

which on integration give, with constants $c_{f,g}$,
\begin{eqnarray*}
f\left( x\right) &=&-\frac{1}{2}\left( 1+\lambda\right) \left\{ \frac{
F_{0}}{k^{2}\left( \lambda^{2}-1\right) }\ln \left[ \cosh \left(
kx\right) \right] +c_{1}x+c_{f}\right\} , \\
g\left( x\right) &=&-\frac{1}{2}\left( 1-\lambda\right) \left\{ \frac{
F_{0}}{k^{2}\left( \lambda^{2}-1\right) }\ln \left[ \cosh \left(
kx\right) \right] +c_{1}x+c_{g}\right\} .
\end{eqnarray*}

The solution has a front which travels at the
velocity $v$ of the source, with also a back front moving at $- c$ (the component moving at $+c$ of course lags behind the superluminal front).

An important point to note is that in the previous solution it was
considered that the source is not spatially limited, so that the component $\Psi_{p}$ ``follows" the source all along. What would happen in the case that the
source is localized in a spatial region, let us say in $0<x<d$ ?

To answer this we consider the equation
\begin{equation}
\frac{1}{c^{2}}\frac{\partial ^{2}\Psi }{\partial t^{2}}-\frac{\partial
^{2}\Psi }{\partial x^{2}}=F\left( x-vt\right) \left[ \Theta \left( x\right)
-\Theta \left( x-d\right) \right] \equiv w\left( x,t\right) ,
\label{Psilocal}
\end{equation}
where $\Theta $ is Heaviside step function.

The general solution of (\ref{Psilocal}) with initial conditions (\ref
{initialc}) is given by
\begin{eqnarray*}
\Psi \left( x,t\right) &=&c\int_{0}^{t}u\left( x,t-s;s\right) ds, \\
u\left( x,t;s\right) &=&\frac{1}{2}\int_{x-ct}^{x+ct}w\left( y,s\right) dy.
\end{eqnarray*}

As an example that can be analytically handled we take again $F=F_{0}\text{sech}^{2}\left( k\xi \right) $. This allows to directly obtain
\begin{eqnarray}
u\left( x,t;s\right) &=&\frac{F_{0}}{2k}\tanh \left[ k\left( x+ct-vs\right) 
\right] \left[ \Theta \left( x+ct\right) -\Theta \left( x+ct-d\right) \right]
\notag \\
&&-\frac{F_{0}}{2k}\tanh \left[ k\left( x-ct-vs\right) \right] \left[ \Theta
\left( x-ct\right) -\Theta \left( x-ct-d\right) \right]  \notag \\
&&+\frac{F_{0}}{2k}\tanh \left[ k\left( d-vs\right) \right] \left[ \Theta
\left( x+ct-d\right) -\Theta \left( x-ct-d\right) \right]  \notag \\
&&+\frac{F_{0}}{2k}\tanh \left( kvs\right) \left[ \Theta \left( x+ct\right)
-\Theta \left( x-ct\right) \right] .  \label{uxts}
\end{eqnarray}

The final solution thus reduces to the sum of eight expressions of the general form
\begin{gather*}
\int_{0}^{t}\tanh \left( \alpha +\beta s\right) \Theta \left( a+bs\right) ds=
\frac{1}{\beta }\ln \left[ \cosh \left( \alpha +\beta t\right) \right]
\Theta \left( a+bt\right) \\
-\frac{1}{\beta }\ln \left[ \cosh \left( \alpha \right) \right] \Theta
\left( a\right) -\frac{1}{\beta }\ln \left[ \cosh \left( \alpha -\frac{a}{b}
\beta \right) \right] \frac{b}{\left\vert b\right\vert }\left[ \Theta \left( 
\frac{a}{b}+t\right) -\Theta \left( \frac{a}{b}\right) \right] ,
\end{gather*}
where the constants $a$, $b$, $\alpha $ and $\beta $ are determined in each
expression taking into account that in (\ref{uxts}) one must make the substitution 
$t\rightarrow t-s$ prior to the integration in $s$.

As can be readily verified, the component of the solution that moves with $v$
is limited to the region $0<x<d$, the components outside this region move
with $\pm c$.

\subsubsection{Einstein equations}
\label{sec:exact}

We now consider a simplified version of the equations in \cite{minotti_modanese_mpla2025} in which the linearization is done on the
scalar fields only, while the geometry of spacetime is treated exactly. We
thus study the Einstein equations derived in \cite{minotti_modanese_mpla2025} with sources linearized in the scalars
\begin{equation*}
R_{\mu \nu }-\frac{1}{2}Rg_{\mu \nu }=\nabla _{\mu }\nabla _{\nu }\phi
-\nabla ^{\alpha }\nabla _{\alpha }\phi g_{\mu \nu },
\end{equation*}
where $\nabla _{\mu }$ represents the covariant derivative. By contraction
with $g^{\mu \nu }$\ to determine $R$, these equations can be written as
\begin{equation}
R_{\mu \nu }=\nabla _{\mu }\nabla _{\nu }\phi +\frac{1}{2}\nabla ^{\alpha
}\nabla _{\alpha }\phi g_{\mu \nu }.  \label{Einstein}
\end{equation}

We now study the possibility of superluminal 1-D solutions for the metric
when the source is not superluminal. For this, considering the expression of
the metric tensor found in the WFA, we take the metric tensor to be of the
form
\begin{equation}
g_{\mu \nu }=
\begin{pmatrix}
g_{00}\left( \xi \right)  & 0 & 0 & g_{03}\left( \xi \right)  \\ 
0 & g_{11}\left( \xi \right)  & 0 & 0 \\ 
0 & 0 & g_{11}\left( \xi \right)  & 0 \\ 
g_{03}\left( \xi \right)  & 0 & 0 & g_{33}\left( \xi \right) 
\end{pmatrix}
,  \label{metric}
\end{equation}
with $\xi =x^{3}-\lambda x^{0}$.

With the assumption that the source to the metric cannot propagate superluminally in vacuum, we take $\phi =0$ in the regions considered. In this case, the equations (\ref{Einstein}) reduce to
\begin{equation}
R_{ik}=0.  \label{Eeqvac}
\end{equation}

The explicit expressions of the ten equations (\ref{Eeqvac}) for the metric (\ref{metric}) result in four non-identically-zero equations corresponding to $R_{00}$, $R_{03}$, $R_{11}$, and $R_{33}$ (we note that $R_{22}=R_{11}$). These equations are very complex, but are linear in the second derivatives of the tensor components.
In this way, one can study the conditions for this system to have algebraic
solutions for the second derivatives of  $
g_{00}\left( \xi \right) $, $g_{03}\left( \xi \right) $, $g_{11}\left( \xi \right) $ and $g_{33}\left(\xi \right) $. 
Thus, considering the inhomogeneous linear system for these derivatives one has that the rank of the matrix of coefficients is two, while that of the augmented matrix is three, indicating that no solution exists. However, the condition that $
g_{11}\left( \xi \right) $ is constant makes the rank of both matrices for the second derivatives of the rest of the tensor components to be equal to one, thus leading to a single independent equation with infinite many solutions. This equation can be conveniently written as (primes indicate
derivatives with respect to $\xi $)
\begin{equation}
4G\left( \xi \right) F^{\prime \prime }\left( \xi \right) -3G^{\prime }\left(
\xi \right) F^{\prime }\left( \xi \right) =0,  \label{g33pp}
\end{equation}
where
\begin{eqnarray*}
F\left( \xi \right)  &\equiv &g_{00}\left( \xi \right) +2\lambda
g_{03}\left( \xi \right) +\lambda ^{2}g_{33}\left( \xi \right) , \\
G\left( \xi \right)  &\equiv &g_{03}^{2}\left( \xi \right)-g_{00}\left( \xi \right) g_{33}\left( \xi
\right)  .
\end{eqnarray*}

Equation (\ref{g33pp}) has the first integral
\begin{equation*}
F^{\prime 4/3}\left( \xi \right) =\kappa G\left( \xi \right) ,
\end{equation*}
with $\kappa $ an arbitrary constant. 

The solutions we are looking for differ from the Minkowski metric in a
localized region around $\xi =0$, so that for $\xi \rightarrow \pm \infty $
they must be constant, which requires that $\kappa =0$. Taking thus into account
the values taken at $\xi \rightarrow \pm \infty $, the resulting metric is
then given by $g_{11}\left( \xi \right) =-1$ and
\begin{equation}
g_{00}\left( \xi \right) +2\lambda g_{03}\left( \xi \right) +\lambda
^{2}g_{33}\left( \xi \right) =1-\lambda ^{2}.  \label{gs}
\end{equation}

It is clear from this relation that given generic expressions for two of the
tensor components, with the only condition that they differ from their
Minkowski values in a localized region around $\xi =0$, the third component
does also satisfy that condition.

In this way, we see that, at variance with the linear wave equation, the
highly non-linear equations (\ref{Eeqvac}) are compatible with
self-sustained superluminal localized solutions. We can thus expect that
once the metric acquires a superluminal component, generated in a region
where an effectively superluminal source acts, this component could persist
traveling in vacuum without need of an exotic source like that required in the ``warp drive" spacetime kind of solutions considered in the literature.

In order to evaluate the effect on an initially stationary mass, we use the geodesic equation describing the neutral mass motion
\begin{equation*}
\frac{d^{2}x^{\mu }}{ds^{2}}+\Gamma _{\alpha \beta }^{\mu }\frac{dx^{\alpha }
}{ds}\frac{dx^{\beta }}{ds}=0.
\end{equation*}

Since we expect that the spatial displacement of the mass during the passage
of the non-flat region is extremely small as compared with the
corresponding $\triangle x^{0}$, we can approximate $ds=c\sqrt{g_{00}}dt$ so
as to write the spatial components of the geodesic equation as
\begin{equation*}
\frac{d^{2}x^{i}}{dt^{2}}=\frac{1}{2}\frac{d\ln \left( g_{00}\right) }{dt}
\frac{dx^{i}}{dt}-\Gamma _{\alpha \beta }^{i}\frac{dx^{\alpha }}{dt}\frac{
dx^{\beta }}{dt}\simeq -c^{2}\Gamma _{00}^{i},
\end{equation*}
where the last expression is justified because the attained velocities of the mass are much smaller than $c$.

We thus have that the mass acquires a specific momentum, after the passage
of the region in the direction of its propagation (since $\Gamma^{1}_{00}=\Gamma^{2}_{00}=0$ there are no lateral deviations), given by
\begin{eqnarray*}
u &=&-c^{2}\int_{-\infty }^{\infty }\Gamma _{00}^{3}dt\simeq-\frac{c}{\lambda }
\int_{-\infty }^{\infty }\Gamma _{00}^{3}d\xi  \\
&=&\frac{c}{2}\int_{-\infty }^{\infty }\frac{g_{03}g_{00}^{\prime }+\lambda
g_{00}g_{33}^{\prime }}{g_{03}^{2}-g_{00}g_{33}}d\xi ,
\end{eqnarray*}
where we have used again the slow motion of the mass to approximate $d\xi=dz-vdt\simeq-vdt$, together with the exact expression of $\Gamma _{00}^{3}$ and relation (\ref{gs}). We note in particular that for the existence of a net effect a non-zero $g_{03}$ is required, since if $g_{03}=0$ a direct integration gives 
\begin{equation*}
u=-\lambda \frac{c}{2}\left. \ln \left[ g_{33}\left( \xi \right) \right]
\right\vert _{\xi =-\infty }^{\xi =\infty }=0.
\end{equation*}

In this scenario the physical source of the effect is primarily the current in the superconductor that gives rise to the e.m.\ scalar $S$, which in turn acts as source of the gravitational scalars $\phi$ and $\psi$. The scalar $\phi$ acts finally as an effective energy-momentum tensor for the spacetime metric, equation (\ref{Einstein}). The superconductor used by Podkletnov has an ordered crystal structure in which the conduction in the $c$ direction, along which the current flows, takes place only through tunneling between $ab$ planes. A possibility is that the electrons tunneling across the length of the superconductor act as an effectively superluminal source \cite{eckle_2008,yu2022} that, as in the example of the wave equation with a superluminal source, eq.\ (\ref{wavesuper}), generates a superluminal field $S$, giving rise to a superluminal scalar $\phi$, resulting ultimately in a superluminal component of the metric tensor. Once generated inside the superconductor, this component could travel superluminally in vacuum as exemplified by the solution (\ref{gs}). This provides a possible interpretation of the lack of a far propagating force in Poher experiments, where a sintered superconductor is used, without ordered crystal structure that could give rise to a long-range, oriented tunneling of electrons.  

\section{Superconducting resonant cavities}
\label{cavities}

In this section we include a brief analysis of another experiment in which the use of superconductors can have an important impact: the generation of thrust in electromagnetic, asymmetric resonant cavities. In  \cite{minotti_modanese_mpla2025} we studied possible thrust
generation in non-superconducting cavities. In that work the derivations were
done using the so called $\gamma $-model \cite{minotti_modanese_epjc2023} to determine the e.m.\ scalar. 

Considering now the case of superconducting cavities, we recall that, as derived in \cite{minotti_modanese_ijmpa2025}, the equation for the e.m.\ scalar $S$ in a superconductor is given by (\ref{SMPLA}), which we restate here:
\begin{equation}
\frac{1}{c^{2}}\frac{\partial ^{2}S}{\partial t^{2}}-\nabla ^{2}S=-\frac{\mu
_{0}q^{2}}{m}\nabla \cdot \left( \left\vert \psi_{bos} \right\vert ^{2}\mathbf{A}
\right) ,  \label{DalaS}
\end{equation}
with
\begin{equation}
\frac{1}{c^{2}}\frac{\partial ^{2}\mathbf{A}}{\partial t^{2}}-\nabla ^{2}
\mathbf{A}+\lambda _{L}^{-2}\mathbf{A}=\mu _{0}\mathbf{j}_{c},  \label{DalaA}
\end{equation}
where the conserved current has the expression 
\begin{equation}
\mathbf{j}_{c}=-\frac{i\hbar q}{2m}\left( \psi_{bos} ^{\ast }\nabla \psi_{bos} -\psi_{bos}
\nabla \psi_{bos} ^{\ast }\right) -\frac{q^{2}}{m}\left\vert \psi_{bos} \right\vert ^{2}
\mathbf{A}.  \label{Jc}
\end{equation}

On the other hand, the equation for the e.m.\ scalar according to the $\gamma $-model used in \cite{minotti_modanese_mpla2025} is
\begin{equation*}
\frac{1}{c^{2}}\frac{\partial ^{2}S}{\partial t^{2}}-\nabla ^{2}S=-\mu
_{0}\gamma \nabla \cdot \mathbf{j}_{c},
\end{equation*}
instead of (\ref{DalaS}).

In this way, if in the expression (\ref{Jc}) the term $\frac{q^{2}}{m}
\left\vert \psi_{bos} \right\vert ^{2}\mathbf{A}$ were to be dominant over the
first term in the right-hand side, eq.\ (\ref{DalaS}) would be equivalent to a $\gamma 
$-model with $\gamma =-1$, and the derivations in \cite{minotti_modanese_mpla2025} would apply
directly to the superconducting cavity.
Thus, the advantage of using a superconducting cavity is that the magnitude of the effective 
$\gamma$ value is much larger than the one expected in ordinary conductors \cite{minotti_modanese_math13050892}.

We must thus compare the terms
\begin{eqnarray*}
T_{1} &=&\frac{i\hbar q}{2m}\left( \psi_{bos} ^{\ast }\nabla \psi_{bos} -\psi_{bos} \nabla
\psi_{bos} ^{\ast }\right) , \\
T_{2} &=&\frac{q^{2}}{m}\left\vert \psi_{bos} \right\vert ^{2}\mathbf{A}.
\end{eqnarray*}

Taking $L$ to be a characteristic scale for spatial variation we thus have
\begin{equation*}
\frac{T_{2}}{T_{1}}\sim \frac{qAL}{\hbar }.
\end{equation*}

On the other hand, from eq.\ (\ref{DalaA}) we can estimate
\begin{equation*}
A\left( \frac{1}{L^{2}}+\frac{1}{\lambda _{L}^{2}}\right) \sim \mu _{0}j_{c},
\end{equation*}
where the parenthesis in the left-hand side takes care of which of the scales is the
important one for the estimation of $A$. We thus have
\begin{equation*}
\frac{T_{2}}{T_{1}}\sim \frac{q\mu _{0}j_{c}}{\hbar }\frac{L^{3}\lambda
_{L}^{2}}{L^{2}+\lambda _{L}^{2}}.
\end{equation*}

The surface charge density in the superconductor is $\sigma \sim \varepsilon
_{0}E_{0}$, with $E_{0}$ the normal electric field at the surface. Since $
j_{c}$ circulates in a sheath of width $\sim \lambda _{L}$ at the surface,
charge continuity imposes that ($\omega $ is the angular frequency of the
considered mode of the cavity)
\begin{equation*}
\omega \sigma \sim \omega \varepsilon _{0}E_{0}\sim \frac{\omega }{c}
j_{c}\lambda _{L},
\end{equation*}
which gives
\begin{equation*}
\frac{T_{2}}{T_{1}}\sim \frac{q\mu _{0}\varepsilon _{0}cE_{0}}{\hbar }\frac{
L^{3}\lambda _{L}}{L^{2}+\lambda _{L}^{2}}=\frac{qE_{0}}{\hbar c}\frac{
L^{3}\lambda _{L}}{L^{2}+\lambda _{L}^{2}}.
\end{equation*}

For microwave cavities the spatial scale imposed by the electromagnetic mode
of frequency $f$, is $L\sim c/f$, typically in the cm range, much larger
than $\lambda _{L}$, a few hundred nm, so that
\begin{equation*}
\frac{T_{2}}{T_{1}}\sim \frac{qE_{0}L\lambda _{L}}{\hbar c}.
\end{equation*}

If we take conservative values $L\simeq 1$ cm, $\lambda _{L}\simeq 400$ nm, $
E_{0}\simeq 1$ kVm$^{-1}$, we obtain $T_{2}/T_{1}\sim 20$, and can thus
apply the derivations in \cite{minotti_modanese_mpla2025} with $\gamma =-1$. 

For instance, for a cylindrical cavity and its TM$_{010}$ mode considered
there we have
\begin{equation*}
E_{0}=\sqrt{\frac{QP}{c\varepsilon _{0}ah}},
\end{equation*}
where $Q$ is the cavity quality factor, $P$ the applied power, $a$ the
radius of the cylinder, and $h$ its height.

In \cite{neunzig2022} the authors tested precisely a cylindrical cavity with superconducting base and top surfaces, with dimensions $a=80$
mm, $h=210$ mm. Unfortunately, the results in \cite{minotti_modanese_mpla2025} cannot be directly
applied because the tested frequencies were well above the 1435 MHz corresponding
to the TM$_{010}$ mode, and, besides, the cavity used is highly symmetric, so
that no net force is expected in principle (no force is also predicted by
the theory if TE modes are used).

Note however that for typical conditions of the experiments: $f=1959$ MHz, $
Q=2063$, $P=18$ W, one has $E_{0}\simeq 29$ kVm$^{-1}$, $L\simeq 15$ cm, so
that $T_{2}/T_{1}\sim 10^{4}$ !

If the TM$_{010}$ mode was employed, with similar values of power and
quality factor, according to the results in \cite{minotti_modanese_mpla2025} with $\gamma =-1$ we can expect a force per unit of mass at each of the end caps of about (in SI units)
\begin{equation*}
f_{z}\simeq \frac{K_{\phi }QP}{c\varepsilon _{0}h}\simeq 6\times
10^{7}K_{\phi }.
\end{equation*}

Using the previous estimation of $K_{\phi }\simeq 10^{-16}$ A$^{2}$N$^{-2}$
we have $f_{z}\simeq 6\times 10^{-9}$ m s$^{-2}$, extremely small to be
detected in the experiments in \cite{neunzig2022}, even allowing for a strong mass unbalance in the caps.

We note that these results apply as long as the scalar $S$ is not too
strong, so that its equation can be linearized. From the estimations above
we have $S\sim A/\lambda _{L}\sim E_{0}/(\omega \lambda_{L})\sim 10$ N A$^{-1}$m$
^{-1}$, which is small as compared with the value at which non-linear effects are expected, given by relation (\ref{Ssat}), which is about two orders of magnitude higher.

\section{Conclusions}

We have studied an extension of the Einstein--Maxwell framework obtained by coupling a scalar--tensor theory of gravity to Aharonov--Bohm electrodynamics, where the scalar quantity $S=\partial_\mu A^\mu$ is promoted to a physical dynamical field. In this setting, the electromagnetic scalar acts as an effective source for the gravitational sector and, in the weak-field and quasi-static limit, generates an additional contribution to the Newtonian potential proportional to $S^2$. This mechanism is absent in ordinary Maxwell electrodynamics and provides the basic ingredient for the electro-gravitational effects discussed in this work.

Our main application concerns strong pulsed discharges through YBCO superconductors. We have shown that, at normal--superconducting junctions, the field $S$ develops a discontinuity proportional to the vector potential and inversely proportional to the London length. Once gravitational backreaction is included, the coupled nonlinear equations admit a saturation regime in the superconducting bulk. This regime prevents the rapid decay of $S$ and defines a threshold value for its buildup. The threshold condition is a central result of the model, because it connects the microscopic field equations with the onset of macroscopic anomalous forces.

When the model is applied to the discharge experiments reported by Podkletnov and Poher, the corresponding threshold condition and force estimates lead to mutually consistent values of the vacuum expectation value $\psi_0$ of the additional gravitational scalar. Although the experimental configurations differ significantly in voltage, pulse duration, and electrode thickness, the inferred values of $\psi_0$ are of the same order of magnitude, which supports the internal coherence of the proposed interpretation. At the same time, this interpretation requires a very large value of $\psi_0$, suggesting that the present Lagrangian should be regarded as an effective description valid in a restricted phenomenological domain rather than as a complete fundamental theory.

An alternative approach is to see if $\psi_0$ can be expressed in terms of fundamental constants. Recall that $\psi_0$ is dimensionless. Using the constants $G_0$ and $\varepsilon_0$ and a mass scale $m$ of the order of the electron mass, the only dimensionless constant that can be obtained is the large Dirac number $N_D=e^2/(4\pi\varepsilon_0 G_0 m^2)\sim 10^{42}$. It is intriguing to hypothesize that this corresponds to $\psi_0^2$. Furthermore, still using the electron mass, the threshold value $S_{sat}$ of the field $S$ can be expressed in terms of a saturation value  $A_{sat}$ of the potential as $S_{sat}=A_{sat}\lambda_L^{-1}$ (Sect.\ \ref{pulsed-discharges}); if $\psi_0^2=N_D$, it follows that $eA_{sat}\simeq mc$.

We have also investigated time-dependent and traveling solutions of the coupled field equations. In the weak-field approximation, localized cylindrical traveling configurations can be constructed and compared with Alcubierre-type kinematics, but here the relevant source is provided by the scalar sector rather than by exotic negative-energy matter. Going beyond the weak-field regime, we obtained a class of one-dimensional exact vacuum traveling solutions of the Einstein equations. 
These results do not by themselves establish the physical propagation mechanism of the observed far-field effects, but they indicate that the theory admits nontrivial dynamical regimes that deserve further investigation.

For superconducting resonant cavities, the same formalism predicts anomalous forces that are much smaller than those associated with pulsed discharge configurations when the value of $\psi_0$ inferred from the discharge experiments is used. This makes such effects difficult to detect with present experimental sensitivities, at least within the parameter range considered here.

Several limitations remain. First, the comparison with the discharge experiments relies on a quasi-static near-field treatment, whereas the reported phenomenology also involves far propagation in a narrow beam. Second, compatibility with solar-system tests requires a strong suppression of the effective scalar coupling to ordinary matter, as discussed in Appendix \ref{app2}. Third, the large inferred value of $\psi_0$ calls for a deeper theoretical justification, possibly within a more complete scalar-tensor model with dynamical screening or attractor behavior toward GR.

Despite these open issues, the framework proposed here identifies a concrete physical mechanism by which superconducting junctions can generate a dynamical electromagnetic scalar, amplify it through nonlinear coupling to gravity, and produce threshold electro-gravitational effects. Further progress now depends mainly on new experiments capable of independently measuring the threshold behavior, spatial profile, time structure, and propagation properties of the anomalous signal.

\appendix 

\section{The discharge experiments}
\label{app1}

The interaction between gravity and macroscopic quantum systems has long been the focus of theoretical and experimental work, since there are reasons to believe that this interaction can differ from that between gravity and classical systems or incoherent ensembles of atoms and molecules. See for example a recent review in \cite{gallerati2022interaction}. Some authors have suggested that quantum aspects of gravitation, still poorly understood, may play a role in this interaction. The debate is open. In our present theoretical model the gravitational dynamics is essentially classical, although the necessity to introduce scalar fields into General Relativity can be seen as a consequence of gravitational vacuum fluctuations.

Among several experiments which display ``anomalies`` in the interaction between gravity and superconductors we have chosen two involving high voltage discharges through superconducting electrodes reported by E.\  Podkletnov \cite{PodkletnovModanese2001} and C.\ Poher \cite{Poher2011APR}, and one with an asymmetric resonating cavity. This latter experiment, theoretically analyzed in Sect.\ \ref{cavities}, is not among those described in the review \cite{gallerati2022interaction}; actually, in Sect.\  \ref{cavities} we do not even model a real experiment, because most experiments about anomalous forces with asymmetrical cavities do not use superconductors. The theoretically predicted forces are in any case very small, if one uses the value of the VEV $\psi_0$ obtained from the data of the two experiments with high voltage discharges.

In the Podkletnov experiment the discharge voltage is very high (500 -- 2000 kV) and the electric pulse duration is short (typically less than 1 $\mu$s; for a comparison between the Podkletnov and Poher experiments also see \cite{modanese2013APR}). The electrode is thick, (4 -- 8 mm) and the superconducting part has an ordered crystal structure. The electrode also has a non-superconducting layer obtained by subjecting one side of the disk to a different treatment. The role of this non-superconductor layer is not completely clear; it could have the effect of re-distributing surface currents in the bulk and could affect the penetration of magnetic flux lines. Therefore the discharge current encounters on one side of the superconductor an NS (normal-superconducting) junction in which N is not a metal; on the opposite side, the superconductor is in contact with a gaseous plasma at low density. The temperature of the superconductor is definitely lower than T$_c \sim$ 92 K, indeed it is between 50 and 70 K, obtained with Helium cooling. 

The observed anomalous gravitational forces are repulsive and are measured with test masses. If there is a force on the electrode itself (the main effect observed by Poher), this is hidden by the fact that the electrode is rigidly fixed. The value of the anomalous gravitational field that we use for comparison with the theoretical model is the one inferred with the test masses dividing their final velocity by the duration of the current pulse. In this comparison, however, we make a simplification which is not yet fully justified (Sect.\ \ref{approximate-time-dep} contains an attempt in this direction): we consider the quasi-static near-field $\chi$, but we do not consider its observed far propagation in a narrow beam. This beam may have the near-field as its ``initial condition" in space, but far away from the superconductor it must clearly obey a wave equation much different from the equation for $\nabla \chi$.

In the Poher experiment such effects of far propagation are not observed. The anomalous field acts, as mentioned, mainly on the electrode itself and its support bar, and only to a lesser extent on some close accelerometers mechanically decoupled from the electrode. The voltage is much lower than in the Podkletnov experiment (a few kilovolts, typically up to $\sim 4\ \mathrm{kV}$), but the current is of similar magnitude (of order $10^{3}$--$10^{4}\ \mathrm{A}$), allowing the threshold value of $S_{\mathrm{jun}}$ to be reached. The duration of the discharge is longer (30 $\mu$s); the superconducting layer is thinner (in total $\sim 60\,\mu\mathrm{m}$), and according to the model the ensuing smaller volume of the superconductor is compensated by the longer duration, because the generated field is weaker but acts for a longer time. The YBCO material is sintered, rather than oriented and single-crystal-like as in the Podkletnov experiment. Both junctions are of the metal--superconductor type. The operating temperature is just below $T_c$

In Poher's experiment the acceleration of the emitter+support assembly is not measured directly with an accelerometer;
it is  inferred  from a sequence of measurements and basic impulse mechanics.
An emitter is clamped to a support mechanically coupled to a linear alternator used as a momentum sensor.
During a discharge, the alternator produces a peak voltage proportional to the velocity of its moving magnets, hence to the
transmitted momentum. The alternator is calibrated experimentally.
From the measured momentum and the measured pulse duration, Poher computes an average force during the discharge.
For example, for a discharge at $U=2863~\mathrm{V}$ ref.\ \cite{Poher2011APR} gives $P=125\pm5~\mathrm{g\,m/s}$ and $\Delta t\simeq 16~\mu\mathrm{s}$,
hence $F_{\mathrm{avg}}\approx 7.8\times10^3~\mathrm{N}$.
Finally, the average acceleration is obtained from Newton's law using the moving mass $m$ of the emitter/support (matched
to the alternator moving mass for efficient momentum transfer).
In the same example this yields $a_{\mathrm{avg}} \approx 1.56\times10^4~\mathrm{m/s^2}$.

Useful information on replication attempts of the experiments of Podkletnov and Poher is contained in refs. \cite{tajmar2022measurement,junker2012setup}. A measurement of the propagation speed of the anomalous force produced in the Podkletnov experiment is described in \cite{podkletnov2012study}. The speed is apparently superluminal, but since the measurement relies on fast readout from piezoelectric sensors a precise evaluation is not simple. Also in view of this we have investigated possible superluminal traveling solutions in Sect.\ \ref{approximate-time-dep}. For a short review of the relation between superluminal signals and causality with reference to recent experiments see \cite{liberati2002faster}.

\section{Constraints on the coupling $\beta_{mat}$}
\label{app2}

In this paper, we do not discuss the complex cosmological implications of the possible existence of a gravitational scalar field with a large vacuum expectation value. It is known that in general it is very difficult to connect the expectation values of fields that appear at the microscopic level in elementary particle physics with the values of the cosmological constant \cite{Martin2012} and of the densities of dark matter, dark energy, and quintessence that are compatible with the observed late cosmic acceleration and CMB anisotropies \cite{Riess1998AJ,Caldwell1998PRL,Aghanim2020PlanckVI}. Furthermore, these comparisons also depend on the scalar-tensor model used and the couplings present in it, which can vary greatly (for example, with a $R\psi$ coupling instead of a $J\psi$ or with a Lagrangian $f(R)$ instead of the Einstein-Hilbert one). 

The minimalist Lagrangian we propose can be adequate to describe electro-gravitational coupling at the microscopic and laboratory scales. Its validity should be judged on the basis of its predictive capabilities at the experimental level. However, it is most likely only an approximation of a more complex model, necessary for extension to large scales. In the works \cite{mbelek2003five,mbelek2004modelling} the compatibility of a model similar to ours (but not in the electromagnetic sector) with some astrophysical observations known in 2003 was studied. Here we consider an internal consistency condition, based on the contribution of $\psi_0$ to the material sources of the Newtonian potential $\chi$, requiring that at least in normal classical gravitational dynamics at the solar system level such effects should not be observable.

In eq.\ \eqref{potem} for the gravitational potential $\chi$ we display only the scalar e.m.\ source proportional to $S^2$ which is responsible for the anomalous effects. However, the full equation (see \cite{minotti_modanese_mpla2025}) also contains a general-relativistic (GR) source proportional to the mass density $T^{mat}_{00}$ and a non-GR mass source proportional to the trace of the tensor $T^{mat}_{\mu\nu}$, denoted with $T^{mat}$.  This non-GR source includes the Brans-Dicke (BD) contribution and that of the source $J$ of $\psi$. In total we have
\begin{equation}
\square \chi=\frac{\partial ^{2}\phi }{\partial t^{2}}+ \frac{1}{2c^2}K_\phi C_\phi S^{2}+
  \frac{4\pi G_0}{(2\omega_0+3)c^2} \left[ \phi_0 \psi_0 \beta_{mat}'-1 \right] T^{mat}
\label{full-box-chi}
\end{equation}
where $\beta_{mat}'$ denotes for simplicity the derivative $\left. \frac{\partial \beta_{mat}}{\partial\phi}\right\vert _{\phi _{0},\psi _{0}}$.

Experimentally, the non-GR contribution is constrained by data from solar-system tests (see below). In our model, this constraint can be satisfied in two ways:

(1) Supposing that $\omega_0$ is large, at least of the order of $10^4$ (a frequent assumption in BD models) and that $\phi_0 \psi_0 \beta_{mat}'$ is of order 1 or smaller. Since our experimental estimate for $\psi_0$ is very large, this requires in turn that $\beta_{mat}'$ is very small.

(2) Supposing that $\omega_0$ is of order 1 and that the two terms in brackets in eq.\ \eqref{full-box-chi} cancel almost exactly. Again, this imposes a strong condition on $\beta_{mat}'$, but with the advantage of relaxing the condition on $\omega_0$. In some sense, the $\psi$ field would have the effect of ``regularizing" BD theory.

The experimental data discussed in Sect.\ \ref{pulsed-discharges} favor the second option, because only for $\omega_0$ of order 1 can the values of the anomalous acceleration, of $S$ at the junctions and $C_\phi$ match.
This is very different from what happens with the function $\beta_{AB}$ which gives the dependence on $\phi$ (and $\psi$) of the coupling of $\psi$ to the e.m.\ scalar. Namely, we assume that $\left. \frac{\partial \beta_{AB}}{\partial\phi}\right\vert _{\phi _{0},\psi _{0}} $ is of order 1. This is equivalent to assume a linear dependence of $\beta_{AB}$ on $\phi$, close to $\phi_0$ and $\psi_0$, of the form
\begin{equation}
  \beta_{AB} \simeq k_\phi \frac{\phi-\phi_0}{\phi_0}
\end{equation}
with $k_\phi$ of order 1. A possible justification for this difference is that while the gravitational scalars must be in some sense in equilibrium with matter, at least in the current cosmological epoch, they cannot be in equilibrium with the e.m.\ scalar, since it is not present anywhere at the macroscopic level.

This argument can be related to the concept of scalar-tensor ``attractor to GR'', based on the fact that local (solar-system) tests constrain an effective scalar-matter coupling, not $\omega_0$ alone.
In the general massless scalar-tensor framework, what high-precision solar-system measurements constrain most directly is the effective coupling of the light scalar degrees of freedom to matter \emph{today}. A  convenient parametrization is obtained in the Einstein-frame formulation, where matter couples to the metric via a conformal factor $A(\varphi)$ and $\varphi$ is a canonically normalized scalar. One defines the (dimensionless) coupling function
$\alpha(\varphi)\;\equiv\;\frac{d\ln A(\varphi)}{d\varphi}$
and its present value $\alpha_0 \equiv \alpha(\varphi_0)$.
The PPN parameter $\gamma$ then takes the form
$\gamma - 1 = -\frac{2\alpha_0^2}{1+\alpha_0^2}$
so a Cassini-type bound on $\gamma-1$ translates into a tight upper bound on $\alpha_0^2$.
A classic experimental determination of $\gamma$ using radio links with the Cassini spacecraft is reported in \cite{bertotti2003test}. General reviews of PPN tests and scalar-tensor frameworks are given in \cite{will2014confrontation}.

If one goes beyond constant-$\omega$ BD theory (e.g.\ allowing $\omega=\omega(\phi)$, additional scalars,
and/or non-minimal couplings), the relation above generally no longer holds: the observable coupling
$\alpha_0$ can depend on additional functions and parameters, and $\omega_0\equiv \omega(\phi_0)$ by itself
does not uniquely determine solar-system deviations. This is what happens in our model, where a value of $\omega_0$ close to 1 can be compensated by the behavior of $\beta_{mat}(\phi,\psi)$.
Damour and Nordtvedt showed that broad classes of tensor--scalar cosmologies generically possess a
dynamical relaxation mechanism toward general relativity during cosmic evolution, in the sense that the
effective coupling $\alpha(\varphi)$ is driven toward zero during the matter-dominated era
\cite{damour1993general,damour1993tensor}.

A closely related idea arises in string-motivated dilaton scenarios: Damour and Polyakov argued that
loop-corrected coupling functions can lead cosmological evolution to drive a (nearly) massless dilaton
toward values where it decouples from matter,
again implying $\alpha_0$ small today and correspondingly small weak-field deviations from GR
\cite{damour1994string}.
In order to apply these ideas, one should of course verify, for the specific model,
that an attractor/decoupling point exists and is stable, and
that cosmological evolution drives the system close enough to it that $\alpha_0^2$ satisfies current bounds. A canonical reduction for dilatonic gravity in 3+1 dimensions has been recently proposed in \cite{scott2016canonical}.

\bibliography{MST_exp} 
\bibliographystyle{ieeetr}

\end{document}